\documentclass[prl, twocolumn,letterpaper,showpacs]{revtex4}
\usepackage{amsmath}
\usepackage{graphicx}

\begin{document}
\begin{abstract}
We analyze the phase conjugate coupling of a pair of optomechanical oscillator modes driven by the time-dependent beat-note of a two-color optical field. The dynamics of the direct and phase conjugate modes exhibit familiar time-reversed qualities, leading to opposite sign temperatures for the modes in the classical regime of operation. These features are limited by quantum effects due to the non-commutative nature of quantum mechanical operators. The effects are measurable by read-out of the oscillator via a qubit. As a potential application of this system in sensing, we discuss a protocol applying phase-conjugate swaps to cancel or reduce external forces on the system.
\end{abstract}
\title{Phase Conjugation in Quantum Optomechanics}
\author{L.~F.~Buchmann, E.~M.~Wright and P.~Meystre}
\affiliation{Department of Physics, College of Optical Sciences and B2 Institute, University of Arizona, Tucson, AZ 85721}

\maketitle

Following impressive successes in cooling optomechanical systems near their quantum ground state~\cite{cooling1, cooling2, cooling3, cooling4}, recent advances have resulted in the demonstration of a number of quantum mechanical effects on a variety of micro- and nanomechanical and ultracold atom platforms~\cite{quantumeff1, quantumeff2,qomreviews}.  A promising new direction is the theoretical~\cite{multimodetheory1, multimodetheory2, multimodetheory3, multimodetheory4, multimodetheory5} and experimental~\cite{multimodeexp1, multimodeexp2} study of multi-mode effects, both optical and mechanical. Important potential applications of quantum optomechanics include for instance the detection of feeble forces and fields at or near the quantum limit~\cite{qomreviews}, high fidelity quantum state transfer between optical and mechanical modes~\cite{braunstein} or between electromagnetic fields of different frequencies~\cite{hill}, and fundamental studies of decoherence and the quantum/classical interface~\cite{quantclass}. Clearly the detailed understanding and control of the effects of both classical and quantum noise are central to such studies.

A technique of noise reduction that has proven useful in optical applications is phase conjugation, which permits to ``time reverse'' and cancel the effects of phase aberrations in the propagation of optical fields. One of its greatest successes is in astronomy, enabling ground-based telescopes to achieve resolutions comparable to or better than space telescopes. In such situations, phase conjugation is achieved by the use of guide stars and deformable mirrors, but in other applications, phase conjugate signals are generated via nonlinearities such as four-wave mixing~\cite{Boydbook}.

This letter extends the idea of phase conjugation to optomechanics. It proposes a specific coupling scheme that permits a mechanical mode of one oscillator to be the phase conjugate of the mode of a direct oscillator. In contrast to the situation in optics, however, this effect occurs in time rather than space. In the classical regime we find as expected that the dynamics of the direct and phase conjugate modes exhibit the time-reversal property familiar from optics. However this feature is limited by quantum noise, a direct consequence of the non-commutative nature of quantum mechanical operators.  When coupling the optomechanical system to a detector qubit, we find that as a consequence the phase conjugate mode interacts with it with an effective {\it negative} temperature, again with important observable corrections in the quantum regime. Finally, we propose a protocol to improve the performance of optomechanical sensors by canceling or reducing certain effects of external forces.

We consider two mechanical modes optomechanically coupled to a cavity optical field. They could be two modes of the same mechanical element, distinct mechanical oscillators \cite{multimodeexp1}, or ultracold atomic clouds~\cite{multimodeexp2}. This system is described by the Hamiltonian ($\hbar =1$)
\begin{equation}
\label{eq:hamiltonian1}
H=H_{\rm cav}+H_{\rm mech}+H_{\rm OM}+H_{\rm diss},
\end{equation}
with the cavity field Hamiltonian
\begin{equation}
\label{eq:opthamiltonian}
H_{\rm cav}=\omega_c\hat{a}^\dag\hat{a}+\eta(t)\hat{a}^\dag+\eta^*(t)\hat{a},
\end{equation}
the mechanical oscillator Hamiltonian
\begin{equation}
\label{eq:mechhamiltonian}
H_{\rm mech}=\sum_{j=1,2}\omega_j\hat{b}_j^\dag\hat{b}_j,
\end{equation}
and the optomechanical interaction Hamiltonian
\begin{equation}
\label{eq:interacthamiltonian}
H_{\rm OM}=\sum_{j=1,2}g_j\hat{a}^\dag\hat{a}(\hat{b}_j+\hat{b}_j^\dag).
\end{equation}
Dissipation of the two subsystems and their respective baths is captured in $H_{\rm diss}$. Cavity photons and mechanical phonons of mode $j$ are annihilated by the bosonic operators $\hat{a}$ and $\hat{b}_j$ respectively, the respective mode frequencies are $\omega_\mathrm{c}$ and $\omega_j$,  $g_j$ are the optomechanical coupling constants, and $\eta(t)$ is the optical pumping rate.

The dynamics of this system were previously studied~\cite{multimodetheory1} for monochromatic laser driving, in which case it was found that the only resonant interactions involve the exchange of excitations, a situation that does not result in phase conjugation. To achieve that goal we consider instead a two-color optical driving field
\begin{equation}
\label{eq:pump}
\eta(t)=\eta_1e^{-i\omega_\mathrm{L1}t}+\eta_2e^{-i\omega_\mathrm{L2}t},
\end{equation}
where $\eta_j$ are complex driving amplitudes. We proceed by decomposing the intracavity field as the sum of a classical mean field $\alpha(t)=\sum_{j=1,2}\alpha_j\exp(-i\omega_{Lj}t)$ and associated quantum fluctuations, with
\begin{equation}
\alpha_j=\frac{-i\eta_j}{\kappa/2+i(\omega_c-\omega_{Lj})},
\label{eq:alphacoeff}
\end{equation}
where $\kappa$ is the resonator damping rate, and  neglect terms quadratic in the fluctuations. Following the time-dependent unitary transformation
\begin{equation}
\label{eq:displacement}
\hat{U}(t)=\exp[\alpha^*(t)\hat{a}-\alpha(t)\hat{a}^\dag],
\end{equation}
the equation of motion for the cavity fluctuations can be integrated applying a slowly-varying envelope approximation for the mechanical modes~\cite{supplement}. The resulting expression is substituted in the Heisenberg-Langevin equations for the mechanical oscillators, which then read
\begin{equation}
\frac{d \hat{\bf b}}{dt}={\rm M}(t)\cdot\hat{\bf b}+\hat{\bf N}+{\bf D},
\label{eq:eom1}
\end{equation}
where $\hat{\bf b}=(\hat{b}_1, \hat{b}_2,\hat{b}_1^\dag,\hat{b}_2^\dag,)^{\rm T}$. Here
\begin{equation}
\mathbf{D}=\left(
\begin{array}{c}
-ig_1|\alpha|^2\\
-ig_2|\alpha|^2\\
ig_1|\alpha|^2\\
ig_2|\alpha|^2\\
\end{array}
\right)
\end{equation}
describes the classical driving of the oscillators, and
\begin{equation}
 \mathrm{M}=\left(
\begin{array}{cccc}
-i\Omega_1-\frac{\Gamma_1}{2} & \mathcal{T}_1 & \mathcal{P}_1 & \mathcal{C}_1\\
\mathcal{T}_2 & -i\Omega_2-\frac{\Gamma_2}{2} & \mathcal{C}_2 & \mathcal{P}_2\\
 \mathcal{P}_1^*&  \mathcal{C}_1^*&i\Omega_1-\frac{\Gamma_1}{2} & \mathcal{T}_1^*\\
\mathcal{C}_2^*&\mathcal{P}_2^*& \mathcal{T}_2^*&i\Omega_2-\frac{\Gamma_2}{2} \\
\end{array}
\right),
\label{couplingmatrix}
\end{equation}
where $\Omega_j$ and $\Gamma_j$ include the optical spring and optomechanical cold damping effects familiar from single-mode optomechanical cooling. The explicit form of all elements of $\rm M(t)$ can be found in the supplemental material~\cite{supplement}. Finally
\begin{equation}
\hat{\mathbf{N}}=\left(
\begin{array}{c}
\hat \zeta_1+\sqrt{\gamma_1}\hat{\xi}_1\\
\hat \zeta_2 +\sqrt{\gamma_2}\hat{\xi}_2\\
\hat \zeta_1^\dagger+\sqrt{\gamma_1}\hat{\xi}_1^\dag\\
\hat \zeta_2^\dagger +\sqrt{\gamma_2}\hat{\xi}_2^\dag\\
\end{array}
\right),
\end{equation}
where $\gamma_j$ and $\xi_j$ account for the noise associated with the coupling of the mechanical oscillators to their thermal reservoirs and
\begin{equation}
\hat \zeta_j=-ig_j(\alpha^*\hat{f}+\alpha\hat{f}^\dag),
\end{equation}
with
\begin{equation}
\label{eq:fnoiseoperator}
\hat{f}(t)=\sqrt{\kappa}e^{(-\kappa/2-i\omega_c)t}\int_0^t {\rm d}t'e^{(\kappa/2+i\omega_c)t'}\hat{a}_\mathrm{in}(t'),
\end{equation}
describes the coupling of the optical cavity noise to the mechanical oscillators, with $\hat a_{\rm in}$ the usual input field noise operator familiar from the input-output formalism~\cite{Wallsbook}.

All elements of the matrix $\rm M(t)$ include constant contributions and terms oscillating at the beat frequency of the two driving fields. Their explicit forms are cumbersome and are relegated to the supplemental material~\cite{supplement}. The important point is that due to their temporal dependence it is possible to choose the frequencies $\omega_{Lj}$ so as to favor specific coupling coefficients, for example the single-mode parametric amplification described by the coefficients $\mathcal{P}_j(t)$, or other forms of mode coupling described by $\mathcal{T}_j(t)$ and $\mathcal{C}_j(t)$, in particular phase conjugation.

Specifically, for the resonant interaction
\begin{equation}
\omega_{L2}-\omega_{L1}=\Omega_1+\Omega_2,
\end{equation}
a condition that is a temporal analog of ``quasi-phase-matching'' in nonlinear optics \cite{Boydbook}, the dominant mode coupling is mediated by the amplitude $\mathcal{C}_j^{+}$ appearing in the oscillatory portion of $\mathcal{C}_j(t)$, see Eq. (18) of the supplementary material. This choice renders all other couplings and interactions, and also the classical driving terms, off-resonant and negligible for sufficiently large separation of $\Omega_1$ and $\Omega_2$, see~\cite{supplement} for explicit expressions. In this regime we can then reduce Eq. (\ref{eq:eom1}) to a $(2\times 2)$ system and the coupling matrix acting on $(\hat b_1, \hat b_2^\dagger)^{\rm T}$ becomes
\begin{equation}
\mathrm{M}_2(t)=
\left(
\begin{array}{cc}
-i\Omega_1-\frac{\Gamma_1}{2} & \mathcal{C}_1^{+}e^{-i(\Omega_1+\Omega_2)t} \\
(\mathcal{C}_2^{+})^*e^{i(\Omega_1+\Omega_2)t}  & i\Omega_2-\frac{\Gamma_2}{2}
\end{array}
\right).
\end{equation}
The nature of the coupling between the mechanical modes is determined by the eigenvalues of $\mathrm{M}_2$. Moving to co-rotating frames and in the limit of negligible dissipation, they are given by  $\pm\sqrt{\mathcal{C}_1^{+}(\mathcal{C}_2^{+})^*}$. If this product is real, we find ourselves in the situation of two-mode parametric amplification, with both modes experiencing gain and becoming entangled~\cite{Boydbook}. Here we focus instead on the case when the eigenvalues are imaginary and the interaction between $\hat{b}_1$ and $\hat{b}_2^\dag$ is oscillatory in nature, i.e. phase conjugation. Unlike two-mode parametric amplification, this evolution associated with $M_2$ alone is not unitary in that case. It is therefore accompanied by quantum noise entering the system through the driving field, the consequences of which will be discussed later.  

The eigenvalues can be tuned by the choice of frequency of the driving lasers. Pure phase conjugation is realized for
\begin{subequations}
\label{eq:detunings}
\begin{eqnarray}
\omega_{L1}&=&\omega_c-\frac{\Omega_1+\Omega_2}{2}\pm\frac{\sqrt{(\Omega_1-\Omega_2)^2-\kappa^2}}{2},\\
\omega_{L2}&=&\omega_c+\frac{\Omega_1+\Omega_2}{2}\pm\frac{\sqrt{(\Omega_1-\Omega_2)^2-\kappa^2}}{2},
\end{eqnarray}
\end{subequations}
which requires that $ |\Omega_1-\Omega_2|>\kappa$. We then have
\begin{subequations}
\label{couplingstrengths}
\begin{align}
\mathcal{C}_1^+&=\frac{i\alpha_1^*\alpha_2g_1g_2(\Delta_1+\Delta_2)}{(\kappa/2-i(\Delta_1-\Omega_2))(\kappa/2+i(\Delta_2+\Omega_2))},\\
(\mathcal{C}_2^+)^*&=\frac{-i\alpha_1\alpha_2^*g_1g_2(\Delta_1+\Delta_2)}{(\kappa/2+i(\Delta_1-\Omega_1))(\kappa/2-i(\Delta_2+\Omega_1))},
\end{align}
\end{subequations}
with $\Delta_j=\omega_c-\omega_{Lj}$ and
\begin{equation}
\mathcal{C}_1^+(\mathcal{C}_2^+)^*=-4|\alpha_1\alpha_2g_1g_2|^2\frac{|(\Omega_1-\Omega_2)^2-\kappa^2|}{\kappa^2(\Omega_1-\Omega_2)^2}.
\label{eq:couplprod}
\end{equation}
While the dynamics of the oscillators do exhibit  phase conjugation effects outside that regime, the dominant contribution in that case is normally parametric amplification. This difference between the oscillator frequencies also justifies the rotating wave approximation made earlier and ensures stability.

To further discuss the properties of optomechanical phase conjugation we absorb free phases into operators, resulting in a real, positive coupling constant $C=|\mathcal{C}_1^+|$. The coupled-mode equations for the mechanical oscillators simplify then to
\begin{subequations}
\label{eq:eom}
\begin{align}
\frac{d\hat{b}_1}{d t}=&(-i\Omega_1-\frac{\Gamma_1}{2})\hat{b}_1+Ce^{-i(\Omega_1+\Omega_2)t}\hat{b}_2^\dag+\hat{\mathcal{F}}_1(t)\\
\frac{d\hat{b}^\dag_2}{d t}=&(i\Omega_2-\frac{\Gamma_2}{2})\hat{b}_2^\dag-Ce^{i(\Omega_1+\Omega_2)t}\hat{b}_1+\hat{\mathcal{F}}_2^\dag(t),
\end{align}
\end{subequations}
where we have combined any external force $\hat F(t)$ applied to the system and noise contributions in the operators
\begin{equation}
\hat{\mathcal{F}}_j(t)=i\hat{F}(t)+\sqrt{\gamma_j}\hat{\xi}_j(t)+\hat{\zeta}_j(t).
\label{sourceterms}
\end{equation}
In what follows we take mode 1 to be the phase conjugate mode, and mode 2 the direct mode.  

Equations~(\ref{eq:eom}) are easily solved in Fourier space to give
\begin{equation}
\tilde{b}_j(\omega)= R_j(\omega)\left [\tilde{\mathcal{F}}_j(\omega)-{\cal L}(\Gamma_k,\omega+\Omega_j)\tilde{\mathcal{F}}_k^\dag(-\omega-\Omega_1-\Omega_2)\right ]
\end{equation}
with $j\neq k$,
\begin{equation}
R_j(\omega)=\frac{1}{\Gamma_j/2+i(\omega+\Omega_j)+C{\cal L}(\Gamma_k,\omega+\Omega_j)},
\end{equation}
and ${\cal L}(\gamma,\omega)=C(\gamma/2+i\omega)^{-1}$. 

The intrinsic noise contribution $\sqrt{\gamma_j}\hat{\xi}_j(t)+\hat{\zeta}_j(t)$ forms a noise floor above which we can measure the effects of the external force $\hat F(t)$. Assuming for now that its strength is well above the noise level we have $\hat{\mathcal{F}}_j(t) \approx i\hat{F}(t)$ in Eq.~(\ref{sourceterms}). In that case, the position operators
\begin{equation}
\tilde{x}_j(\omega) \equiv \tilde{b}_j(\omega)+\tilde{b}_j^\dag(-\omega)
\end{equation}
become
\begin{eqnarray}
\tilde{x}_j(\omega)&=&\chi_j(\omega)\tilde{F}(\omega)+\chi_{j,c}(\omega)\tilde{F}(\omega+\Omega_1+\Omega_2) \nonumber\\
&+&\chi_{j,c}^*(-\omega)\tilde{F}(\omega-\Omega_1-\Omega_2),
\label{xofomega}
\end{eqnarray}
with
\begin{eqnarray}
\chi_j(\omega)&=&i(R_j(\omega)-R_j^*(-\omega))\\
\chi_{j,c}(\omega)&=&-iR_j(\omega){\cal L}(\Gamma_k,\omega+\Omega_j).
\end{eqnarray}

It is straightforward to evaluate the spectral densities $S_{xx,j}(\omega)=\int\mathrm{d}\omega'\langle \tilde{x}_j(\omega)\tilde{x}_j(\omega')\rangle$ of the mechanical oscillator positions for the case of a stationary force, for example stationary noise, for which $\langle\tilde{F}(\omega)\tilde{F}(\omega')\rangle=\delta(\omega+\omega')S_F(\omega)$ with $S_F(\omega)=\int\mathrm{d}te^{-i\omega t}\langle \hat{F}(t) \hat{F}(0) \rangle$. Their full analytic expressions are cumbersome and can be found in the supplemental material~\cite{supplement}. For qualitative insight we reproduce only the contributions to $S_{xx,1}(\Omega_1)$ proportional to $S_F(\pm\Omega_2)$. This corresponds to the response of the phase conjugate mode 1 to a narrow-width external force applied to the direct mode 2  only. We find
\begin{eqnarray}
S_{xx,1}(\Omega_1)&\propto& \chi_{1,c}^*(-\Omega_1)\left[\chi_1(\Omega_2)+\chi_{1,c}(-\Omega_1)\right.\nonumber\\
&&\left.+\chi_{1,c}^*(-\Omega_1-2\Omega_2)\right]S_F(-\Omega_2)
\label{eq:sxxpropto}
\end{eqnarray}
The key point here is that the spectral density at the positive frequency $\Omega_1$ depends on the external force noise power spectrum at the {\it negative} frequency, $-\Omega_2$.
This is a signature of the time reversal property of phase conjugation, and is in stark contrast to the familiar situation of linearly coupled oscillators. In the latter case the susceptibilities are the same, but the effect of phase conjugation is to change signs in the arguments of $\tilde{F}(\omega)$ in the last two terms of Eq.~(\ref{xofomega}). (Note that or parametrically amplified oscillators, the gain from the drive prevents thermalization of the system in the absence of saturation.)

\begin{figure}
\includegraphics[width=1\columnwidth]{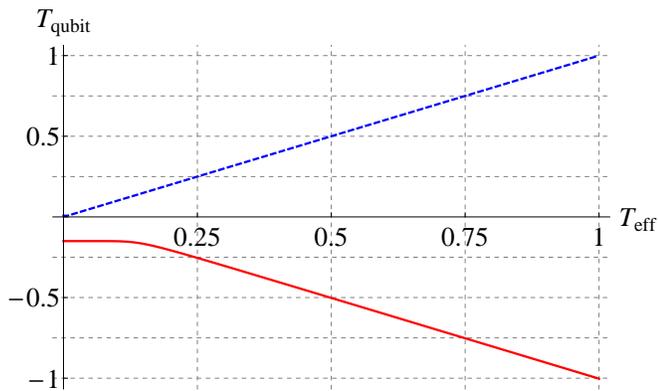}
\caption{Temperature of the qubit for linearly (blue, dashed) and phase conjugate (red, solid) coupled oscillators as a function of force temperature in natural units. The parameters are $\Omega_1=1$, $\Omega_2=1.5$, $\gamma_1=\gamma_2=0.1$, $C=0.025$.}
\label{temperatures}
\end{figure}

Consider for example a force with effective temperature $T_\mathrm{eff}$ acting on the direct mode only, so that~\cite{Clerk}
\begin{equation}
\label{eq:noisetemp}
\frac{S_F(\omega)}{S_F(-\omega)}=\exp\left(\frac{\hbar\omega}{k_BT_\mathrm{eff}(\omega)}\right).
\end{equation}
As a result of the phase conjugate coupling, oscillator 1 experiences then a force with {\it negative} temperature $-T_\mathrm{eff}$.  This has physically observable consequences, which can be seen for example by coupling that oscillator to a qubit of transition frequency $\Omega_1$ and upper to lower level decay rate $\Gamma$ through the interaction Hamiltonian~\cite{Schwab}
\begin{equation}
V= A\hat{x}_1\hat{\sigma}_{x},
\end{equation}
a simple model of a spectrum analyzer. The Fermi Golden Rule transition rates of the qubit are
\begin{eqnarray}
\Gamma_{\mathrm{g}\to\mathrm{e}}&=&A^2S_{xx,1}(-\Omega_1)\\
\Gamma_{\mathrm{e}\to\mathrm{g}}&=&A^2S_{xx,1}(\Omega_1).
\end{eqnarray}
It is straightforward to determine the steady-state occupation of the two states of the qubit, and to infer its temperature as a function of the temperature of the external force.  This situation is depicted in Fig.~\ref{temperatures}, which shows that indeed, for high enough temperatures the qubit equilibrates at $-T_{\rm eff}$.
Importantly, though, the situation is fundamentally different for low temperatures, where the temperature of the qubit levels to a constant value. This is due to the quantum noise terms $\sqrt{\gamma_i}\hat{\xi}_i(t)+\hat{\zeta}_i(t)$ of Eq.~(\ref{sourceterms}), which we have ignored so far in the discussion (but not the numerics) and which limit the effectiveness of phase conjugation in the quantum regime. That these terms should be important is easily understood by recalling that phase conjugation achieves a situation where the expectation values of the annihilation operator of mode 1 and creation operator of mode 2 become equal,
\begin{equation}
\langle \hat b_1\rangle = \langle \hat b_2^\dagger\rangle.
\end{equation}
This equality cannot be valid at the level of operators, since this would violate the boson commutation relations. The fundamental quantum noise present in phase conjugation is essential in preserving them, and the flattening of the negative temperature of the qubit for $T_{\rm eff} \rightarrow 0$ is a direct signature of that noise. This fundamental element of phase conjugation in the quantum regime was first realized in the optical case~\cite{Gaeta}, where it was shown that except in special cases~\cite{Bajer} quantum noise imposes a limitation on the ability of phase conjugation to reverse the effect of external fluctuations below the classical limit.

\begin{figure}
\includegraphics[width=1\columnwidth]{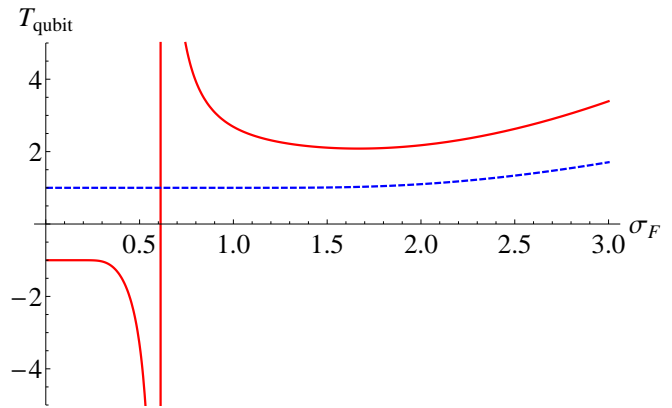}
\caption{Qubit temperature as a function of the width $\sigma_F$ of $S_F(\omega)$ for linearly (blue, dashed) and phase conjugate (red, solid) coupled oscillators. Other parameters as in Fig.~\ref{temperatures}.}
\label{noisewidth}
\end{figure}

We also remark that the situation is more complex if the spectral width $\sigma_F$ of  $S_F(\omega)$ becomes large enough to drive both modes directly. This situation is depicted in Fig.~\ref{noisewidth}: for $\sigma_F \ll |\Omega_1 - \Omega_2|$  oscillator 1 is subject to the time-reversed version of $F(t)$ only. As $\sigma_F$ increases, oscillator 1 starts to experience the direct effect of $F(t)$ as well. This eventually overwhelms the contribution due to phase conjugation, with a relatively sharp transition where the oscillators go through a resonance, but with the temperature remaining finite throughout.

We conclude by outlining a possible protocol that exploits optomechanical phase conjugation to reduce the distortion due to a classical external force $F_j(t)$ affecting the oscillators. Transforming to their respective frames of reference, $\hat{B}_j=e^{i\Omega_j t}\hat{b}_j$, the free evolution is given by
\begin{equation}
\label{eq:timeevol}
\hat{B}_j(t)=\hat{B}_j(t_0)-i\int_{t_0}^t\mathrm{d}t'e^{i\Omega_j t'}F_j(t').
\end{equation}
Starting at $t=-t_1$, both oscillators evolve freely and are subject to their respective forces. At $t=0$, we apply a phase conjugate swap of the two oscillators, fast enough so we can neglect the effect of the force during that process. Following this we let the system evolve freely for a time $t_2$. After these steps the state of the oscillators is
\begin{align}
\label{eq:secondtimeevol}
\hat{B}_j(t_2)=&\hat{B}_{k}^\dag(-t_1)+i\int_{-t_1}^{0}\mathrm{d}t'e^{-i\Omega_k t'}F_k(t')\nonumber\\
&-i\int_{0}^{t_2}\mathrm{d}t'e^{i\Omega_j t'}F_j(t')+\hat{N}_\mathrm{PC,j}\nonumber \\
=&\hat{B}_{k}^\dag(-t_1) + \hat{N}_\mathrm{PC,j} \nonumber \\
&+i\int_{0}^{\tau}\mathrm{d}\tau'e^{-i\tau'}\left [\frac{F_k\left (-\frac{\tau'}{\Omega_k}\right )}{\Omega_k}-\frac{F_j \left (\frac{\tau'}{\Omega_j}\right )}{\Omega_j}\right ], 
\end{align}
where we have used the ``echo condition''
\begin{equation}
\Omega_k t_1 =\Omega_j t_2 \equiv \tau.
\end{equation}
Classical noise cancellation occurs if the term in square brackets vanishes. In optical phase conjugation this is easy to achieve since the degenerate situation $\Omega_k=\Omega_j$ is possible. However, this is not the case here. Specifically for a truly static force, cancelation occurs for $F_k/\Omega_k=F_j/\Omega_j$. More generally for ``quasi-static'' forces that vary slowly compared to the inverse oscillator frequencies we can still expect a significant degree of noise reduction. This is similar to the situation in optics, where for instance astronomical  guide star techniques rely explicitly on the slow rate of change of atmospheric distortions. A more detailed analysis of optomechanical phase-conjugation based noise reduction techniques will be the object of future studies.

In conclusion, we have proposed a scheme to achieve phase conjugate coupling between two optomechanically driven mechanical oscillators.   In a stationary setting, it formally swaps  the roles of emission and absorption of excitations with an external force, leading to negative temperatures that could be measured by coupling a qubit to the position of an oscillator. However, unavoidable quantum noise limits this process in the low-temperature limit where $\hbar \Omega_j$ becomes comparable to the effective thermal energy of the force, $k_B T_{\rm eff}$. Phase conjugate swapping can be used to reduce or cancel effects of external forces acting on the oscillators.

This work is supported by the DARPA QuASAR program through a grant from AFOSR, the DARPA ORCHID program through a grant from ARO, the US Army Research Office, and the National Science Foundation.

\end{document}